# Exciton-polariton condensates near the Dirac point in a triangular lattice

N Y Kim[1§], K Kusudo[2,3], A Löffler[4], S Höfling[4], A Forchel[4] and Y Yamamoto[1,2]

[1]E. L. Ginzton Laboratory, Stanford University, Stanford CA 94305-4085, USA.
[2]National Institute of Informatics, 2-1-2 Hitotsubashi, Chiyoda-ku, Tokyo 101-8430, Japan.
[3]NTT Basic Research Laboratories, 3-1 Morinosata-Wakamiya, Atsugi-shi, Katagawa, 243-0198, Japan.
[4]Technische Physik and Wilhelm-Conrad-Röntgen-Research Center for Complex Material Systems, Universität Würzburg, D-97074 Würzburg, Am Hubland, Germany.

E-mail: nayoung@stanford.edu

**Abstract.** Dirac particles, massless relativistic entities, obey linear energy dispersions and hold important implication in particle physics. Recent discovery of Dirac fermions in condensed matter systems including graphene and topological insulators raises great interests to explore relativistic properties associated with Dirac physics in solid-state materials. In addition, there are stimulating research activities to engineer Dirac paricles to eludicte their physical properties in a controllable setting. One of the successful platforms is the ultracold atom-optical lattice system, whose dynamics can be manipulated in a clean environment. A microcavity exciton-polariton-lattice system provides an alternative route with an advantage of forming high-orbital condensation in non-equilibrium conditions, which enables to explore novel quantum orbital order in two dimensions. Here we directly map the liner dispersions near the Dirac points, the vertices of the first hexagonal Brillouin zone from exciton-polariton condensates trapped in a triangular lattice. The associated velocity values are ~ 0.9 - $2\times10^8$ cm/s, which are consistent with the theoretical estimate $1\times10^8$ cm/s with a 2 $\mu$m-lattice constant. We envision that the exciton-polariton condensates in lattices would be a promising solid-state platform, where the system order parameter can be accesses in both real- and momentum spaces. We furthermore explore unique phenomena revealing quantum bose nature such as superfluidity and distinct features analogous to quantum Hall effect pertinent to time-reversal symmetry.

1. Introduction

In 1928, a brilliant equation dawned on Paul Dirac, who had searched a theory to describe the relativistic motion of electrons combining relativity and quantum mechanics. He reached a massless linear dispersion relation purely mathematically, which preserves the space-time symmetry in the form of the first-order derivates. Dirac equation indeed successfully describes the hydrogen atomic spectrum and predicts a new form of matter such as positron [1] in particle physics. The linear massless dispersion of relativistic Dirac particle is often expressed in a Dirac-Weyl Hamiltonian form as $H = \hbar v \vec{\sigma} \cdot \vec{q}$, where $\hbar$ is the normalized Planck's constant by $2\pi$, $v$ is the Dirac particle velocity, $\vec{\sigma}$ is the set of the Pauli matrices and $\vec{q}$ is the momentum wavevector. The linear Dirac dispersions appear in the band structure of a single particle confined in certain two dimensional lattices. The famous example is the honeycomb lattice, where two atoms in a unit cell act as pseudo-spins represented by the 2×2 Pauli spin matrices. However, the honeycomb geometry is not unique and Dirac particles can emerge in other fundamental 2D lattices such as line-centered square or kagome lattices as long as two degenerate Bloch bands of the periodic potentials are crossing to form a gapless conical singularity as known as a Dirac point. Calculations showed that the band structures of the triangular photonic lattices have the Dirac points formed by two excited bands at the vertices of the first Brillouin zone [2, 3].

This relativistic quantum mechanical equation provides a great insight on the emergence of Dirac fermions in recently discovered two-dimensional condensed matter materials including graphene [4], carbon nanotubes [5, 6], and topological insulators [7, 8], which exhibit linear electronic band structures at the Dirac points. These materials have already shown exotic transport properties, for example, anomalous quantum Hall effect [9], Klein tunnelling [10] and *Zitterbewegung* [11]. Although the aforementioned properties would be attributed to Dirac fermions, it is rather indirect to verify the existence of Dirac fermions. The only direct evidence of Dirac fermions in these systems comes from angle-resolved photoemission spectroscopy, which can map the linear energy-momentum relation and furthermore capture the unique kink structure in the spectra potentially induced by interactions among particles [12]. Despite of excitement and success in identifying the fascinating properties of relativistic Dirac particles in condensed matter materials, disorder, imperfectness and complicated long-range Coulomb interactions in real materials may challenge us to investigate Dirac physics.

In order to overcome challenges and difficulties in real materials, there have been myriads of theoretical proposals to engineer artificial Dirac particles using both fermions and bosons and to mention practical methods for accessing their unique signatures in 2D electronic and photonic platforms: electrons in two-dimensional electron gas systems [13, 14], metal nanoparticles in plasmonic lattices [15], photons in triangular or honeycomb photonic crystals [16-22], two trapped ions [23], atomic condensates in honeycomb optical lattices [24, 25], fermionic atoms in $T_3$ optical lattice [26,27], and fermionic and bosonic atoms in line-centred square optical lattices [28,29] which would have a capability to construct the rich phase diagram governed by Dirac physics. In particular, the superfluidity of Bose-Einstein condensates at the Dirac point in the honeycomb optical lattice would be predicted to leave a marked fingerprint near the Dirac points [30]. Such permanent signature would be robust for atoms with non-zero interactions trapped in any lattices holding a triangular symmetry. This prediction provides an exciting insight that the altered Dirac band structure may be a unique feature of Dirac bosons arising from the interplay of superfluidity of interacting condensates and massless relativistic Dirac motion. As an experimental progress, one-dimensional Dirac equation has been simulated in a single trapped ion [31], and Dirac-point engineering has been also demonstrated in a tunable honeycomb optical lattice [32] and in a manipulated carbon-atom hexagonal array on the copper-oxide surface [33]. However, the direct detection of the band structure in various engineered Dirac systems remains elusive or technically complicated. Here we prepare exciton-polariton condensates in a 2D triangular lattice, where the linear dispersion zones are formed by two degenerate $p$-like orbitals. We directly map the band structures via the standard spectroscopy measurements in the momentum space.

2. Microcavity exciton-polaritons in a triangular lattice

2.1 Microcavity exciton-polariton condensation

Microcavity exciton-polaritons possess light-matter duality inherited from the strongly coupled cavity photons and quantum-well (QW) excitons. In the low density limit, (spinless) exciton-polaritons are indistinguishable quantum Bose particles and undergo a phase transition to condensation, a spontaneous buildup of massive macroscopic population in a single ground state via stimulated scattering among weakly interacting particles. Phase transition occurs around 4 K in inorganic semiconductors such as CdTe [34] and GaAs [35-38] and around room temperatures in organic materials [39] and large-band gap GaN semiconductors [40]. Such elevated critical temperatures in exciton-polariton condensation, around $10^7$-$10^9$ times higher than those of atomic Bose-Einstein condensates, result from extremely light effective mass of exciton-polaritons primarily due to photonic fraction. A nonzero photonic component is also a primary source of the finite lifetime of exciton-polaritons, which naturally leak through the cavity. Capturing leaked photons whose energy and in-plane momentum information reveal those of polaritons (one-to-one correspondence), the extensive evidence of exciton-polariton condensates in the ground state has been found: an exponential increase in the ground state occupancy above quantum degeneracy threshold, narrower distribution in momentum space, increased spontaneous long-range spatial coherence and bunched second-order correlations [34-46].

The spontaneous decay of exciton-polaritons is compensated by continuous injection of particles by external pumping, making exciton-polariton condensates stabilized in non-equilibrium condition distinct from atomic counterparts. This open-dissipative nature of exciton-polariton condensates is advantageous in giving rise to coherent states with high-orbital symmetry beyond the ground state as a consequence of the competition between the relaxation scattering time to the ground state and the short lifetime of the states. When we introduce one-dimensional and 2D periodic potentials in exciton-polariton devices, well-defined meta-stable coherent states are accessed in high symmetry points of given lattice potentials. In this manner, coherent *p*- and *d*-wave symmetry of exciton-polariton condensates have been observed in one-dimensional array [37] and 2D square lattice potentials [47], respectively. This work focuses on exciton-polariton condensates in 2D triangular lattice potentials.

2.2 Two-dimensional triangular lattice band structure calculation

As a simplest example of 2D hexagonal lattice group, the triangular lattice is generated by two non-orthogonal unit vectors in the real space denoted as $\vec{a}_1 = a(1,\ 0)$ and $\vec{a}_2 = a\left(-\frac{1}{2},\ \frac{\sqrt{3}}{2}\right)$, where $a$ is the distance between nearest neighbour sites (Fig. 1(a)). The corresponding reciprocal lattice vector bases $\vec{b}_1 = \frac{2\pi}{a}\left(1,\ -\frac{1}{\sqrt{3}}\right)$ and $\vec{b}_2 = \frac{2\pi}{a}\left(0,\ \frac{2}{\sqrt{3}}\right)$ construct hexagonal Brillouin zones (BZs). The triangular lattice holds three high symmetry points **Γ, *M*** and ***K***, one rotation center of order 6 (**Γ**), three of order 2 (***M***) and two of order 3 (***K***) respectively (Fig. 1(b)). In particular, two inequivalent ***K*** and ***K'*** points in the zone boundary of the first hexagonal BZ are related under the inversion symmetry. This structure shares common rotational symmetry with that of the honeycomb lattice. The general reciprocal lattice vectors $\vec{G}_{mn}$ are spanned by $\vec{b}_1$ and $\vec{b}_2$ as $\vec{G}_{mn} = m\vec{b}_1 + n\vec{b}_2$ with integers *m* and *n*.

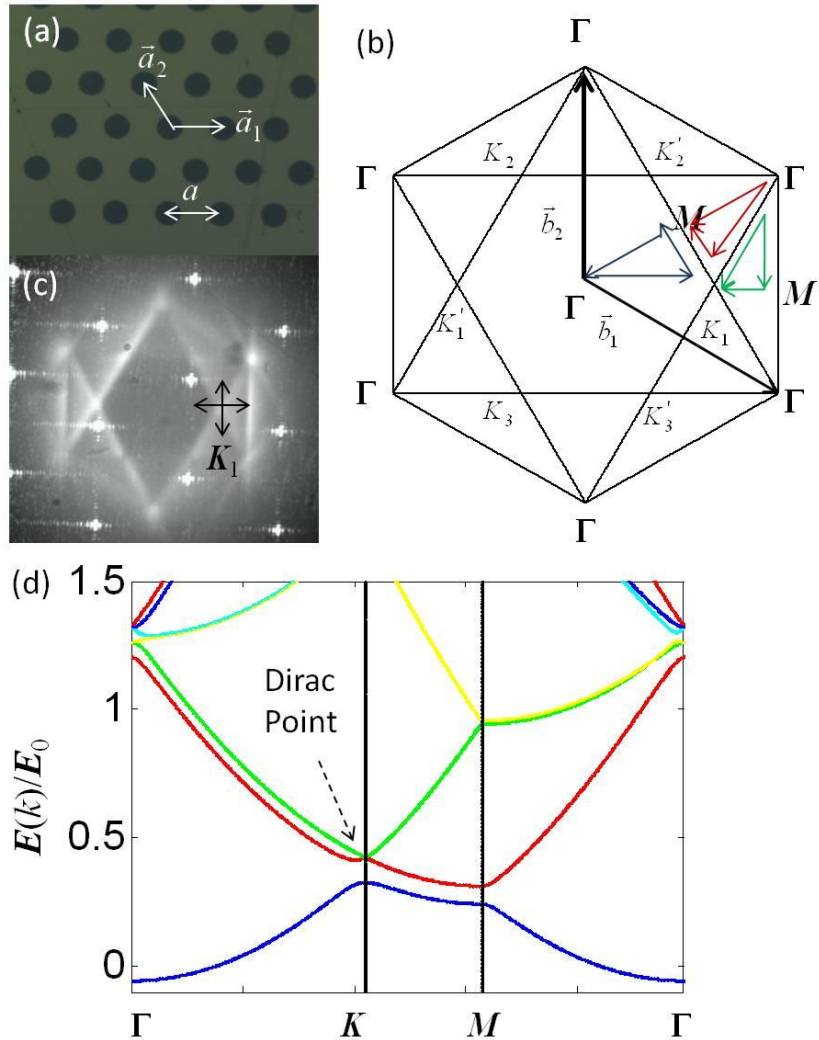

**Figure 1** (a) A photograph of the 2D triangular pattern in a thin metal film on top of the wafer. Brighter regions are where the metal film is deposited and darker ones are bare semiconductor surface. Theoretical (b) and experimental (c) first three hexagonal Brillouin zones (BZs) of the 2D triangular lattice with three high symmetry points $\Gamma$, $M$ and $K$. (b) Individual three $K$ and $K'$ points are labelled with index 1, 2, and 3. Colored triangles along $\Gamma$, $M$ and $K$ for band structures are indicated in blue (the first band), red (the second band) and green (the third band) (d) Calculated single-particle band structures normalized by the characteristic kinetic energy $E_0$ along three high symmetry points $\Gamma$, $K$ and $M$. Dirac point is specified where two degenerate $p$-orbital bands meet.

The single particle band structure is computed by solving the Schrödinger equation with an effective lower polariton (LP) mass $m^*$, $-\dfrac{\hbar^2}{2m^*}\nabla^2\Psi(\vec{r}) + V(\vec{r})\Psi(\vec{r}) = E\Psi(\vec{r}),$ in the plane-wave basis $|\vec{k}+\vec{G}_{mn}\rangle$ within the first Brillouin zone. The triangular periodic potential $V(\vec{r})$ is assumed to have a form, $V(\vec{r}) = \sum_i g(r - \vec{R}_i)$,

$$g(\vec{r}) = \begin{cases} -V_0 & \text{for } r < r_0 \\ 0 & \text{for } r > r_0 \end{cases}$$

, where $\vec{R}_i$ is the location of lattice site center, $g(\vec{r})$ is the potential of circular wells with radius, $r_0$ and the potential strength, $V_0$. The Hamiltonian operator, $\hat{H} = -\frac{\hbar^2}{2m^*}\hat{k}^2 + \hat{V}(x,y)$, is expressed in a matrix form by computing $\langle \vec{k}+\vec{G}_{mn}|\hat{H}|\vec{k}+\vec{G}_{mn}\rangle$ and is diagonalized. The single-particle plane-wave basis method is justified to model our system since the generated potential strength $V_0$ induced by a thin metal film, around 200 μeV, is much weaker than characteristic kinetic energy $E_0 = -\frac{\hbar^2}{2m^*}\left(\frac{2\pi}{a}\right)^2 \sim 4.5$ meV with effective mass $m^* \sim 8.3\times10^{-5} m_0$ ($m_0$ is electron mass) and $a = 2$ μm.

Figure 1(d) plots a representative band structure in reduced-zone scheme along the high symmetry triangular loops in the order of **Γ**, **K**, **M** to **Γ**. The periodic potential lifts energy degeneracy at high symmetry points. At **K** points, three degenerate energy states are split into degenerate doublet and lowest non-degenerate singlet. The second and third bands form six Dirac cones at the vertices of the first BZ, and the conical singularity per vertex, a Dirac point, emerges where degenerate doublets meet.

### 2.3 Dirac dispersion

Near Dirac points at **K** and **K'** ($|\vec{K}| = \frac{2\pi}{a}\left(\frac{2}{3}\right)$, the magnitude of the wavenumber at **K** and **K'**), we can establish the effective Hamiltonian matrix $H_{eff}$ expressed as a 2×2 matrix in terms of degenerate doublet wavefunctions and Pauli matrices, $\sigma_x$ and $\sigma_z$,

$$H_{\text{eff}} = \left(\frac{\hbar^2 k_{//}^2}{2m^*} + \frac{\hbar^2 q^2}{2m^*} + V_0\right)\mathbf{1} - \hbar v(q_x \sigma_z + q_y \sigma_x),$$ where $\mathbf{1}$ is a two-by-two identity matrix and $v$ is the relativistic velocity defined as $v = h/3m^*a$. The corresponding energy eigenvalues of $H_{\text{eff}}$ at $\vec{K} + \vec{q}$ are described by $E(\vec{K} \pm \vec{q}) = E_{\vec{K}} + E_{\vec{q}} + V_0 \pm \hbar v |\vec{q}|$, for a small non-zero momentum deviation $\vec{q}$ away from $K$ points and the kinetic energy, $E_{\vec{K}} = \frac{\hbar^2 \vec{K}^2}{2m^*}$. It is the equation of massless linear Dirac dispersions. For a given structure ($a$ = 2 $\mu$m, $m^* \sim 8.3 \times 10^{-5} m_0$), the velocity is estimated $1.5 \times 10^8$ cm/s $\sim 0.003c$ where $c$ is the speed of light in vacuum.

3. Experiment and Discussion

   3.1 Device and experimental setup

Our microcavity consists of the 16(20) GaAs/AlGaAs mirror-pairs in the top(bottom) distributed Bragg reflector and three stacks of 4 GaAs quantum wells (QWs), which are placed at the central antinodes of the intrinsic AlAs $\lambda/2$ cavity. The measured quality factor ($Q$) of lower polariton (LP) emission spectra below quantum degeneracy threshold is ~ 1800 at detuning value~ - 3 meV, where detuning is defined to be the cavity photon energy offset with respect to the QW exciton energy at zero in-plane momentum. The QW-microcavity structure exhibits ~ 13.8 meV vacuum Rabi splitting near zero detuning. We have patterned a 2D triangular lattice, whose lattice constant $a$ is 2 $\mu$m by a thin-metal film deposition method (23/7 nm Au/Ti film) similar to our previous samples [37, 43, 48] (Fig. 1(a)).

A compact cryo-optical setup for micro-photoluminescence experiments has been constructed both in near-field and far-field spaces utilizing a straightforward Fourier optics technique. This setup has been proved to be powerful to access spatially (near-field) and angularly (far-field) resolved spectroscopy and imaging. The QW-microcavity device is held at ~ 4 K and excited at ~ 60 degree by a 3 ps pulsed laser at 767.7 nm with a repetition rate of 76 MHz. The excited laser wavenumber is $k_{//} \sim 7.39 \times 10^4$ ($\sim 2.4 \times \left(\frac{2\pi}{a}\right)$) cm$^{-1}$, and its energy is 6 meV higher from LP state energy. The LP emission is collected through a 0.55 NA microscope objective lens and the signals are fed into either a 750 mm grating spectrometer whose angular resolution is ~ 0.3° and spectral resolution is ~ 0.02 nm with a nitrogen–cooled CCD for spectroscopy or a CCD camera for imaging.

### 3.2 Experimental Results

#### 3.2.1 Lower polariton condensates near the Dirac points

Figure 2 shows the LP population distributions in the momentum spaces at different pump power values normalized by the threshold power value $P_{th}$. The bright triangular peaks come from the scattered laser signal at 60 degree, with which we set the momentum space unit vector amplitude drawn in the white bar in Fig. 2(a), $|\vec{b}_1| = |\vec{b}_2| = \frac{2\pi}{a}\left(\frac{2}{\sqrt{3}}\right)$. Below threshold (Fig. 2(a)), the emission pattern of LPs is a broad, isotropic donut-shaped, which indicates the bottleneck effect. Elaborately, LPs decay through the cavity from high energy large momentum bands before relaxing to lower energy small momentum states. This is because stimulated scattering induced by repulsive particle-particle interaction is insufficient to reach lower momentum states (**Γ** point). However, as soon as the LP density is above the quantum degeneracy threshold (Fig. 2(b)), the first BZ is clearly seen with developed sharp peaks at **K** and **K'** points as well as the edges of the first BZ. In stark contrast to the boarder isotropic circular shaped momentum distribution, narrower momentum linewidth along the first BZ implies coherence among condensed exciton-polaritons. The more LPs are created at higher pump powers, the LP population inside the first BZ grows (Fig. 2(g)). Note that these time-integrated images record the emission patterns from LPs in all energy values, and the energy-resolved LP population distribution in the momentum space is measured with the spectrometer.

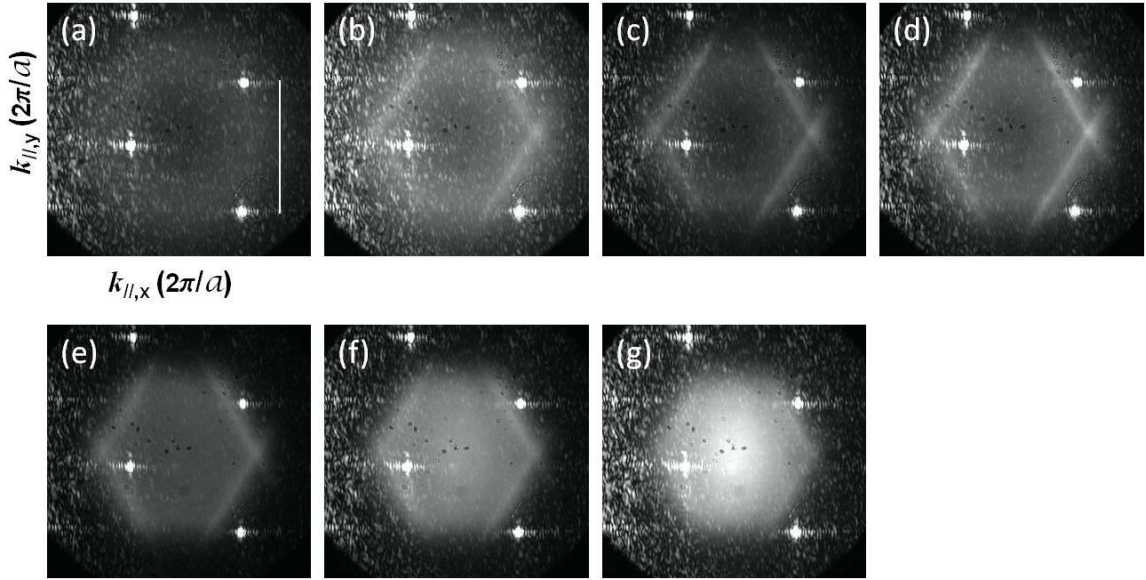

**Figure 2** Lower polariton population distributions in the momentum ($k_{//,x}$, $k_{//,y}$) space as a function of the laser power values $P/P_{th}$ ~ 0.5, 1, 1.5, 2.1, 6.3, 14, 25 from (**a**) to (**g**). The power is normalized by the threshold power value $P_{th}$ ~ 4 mW.

We collect the signatures of condensates at all $K$ and $K'$ points from excitation pump-power dependent spectroscopic measurements. The experimental results provide primarily three physical quantities: (1) maximal photoluminescence signal intensity, which is proportional to the LP population, (2) the energy of LP state, and (3) the spectral linewidth of the LP signal defined as a full-width a half-maximum. The log-log plots of the LP intensities at all $K$ and $K'$ points in Fig. 3(a) are presented against the normalized laser pump power values $P/P_{th}$, where $P_{th}$ ~ 4 mW taken at ~ 4 K. The nonlinear increase in the LP intensity trend above $P/P_{th}$ ~ 1 exist at all $K$ and $K'$ points, indicating the emergence of massive population at these meta-stable points. The intensity saturates and falls down far above pump power values ($P/P_{th} > 7$) because polaritons are finally scattered into the global ground state at the zone-center $\Gamma$ point ($\vec{k} = 0$). Furthermore, the blue-shift in energy below and above threshold shown in Fig. 3(b) and the spectral linewidth reduction near threshold and the broadened spectral linewidth above threshold in Fig. 3(b) have been clearly observed. It is convincing that all three experimental features near Dirac points, which share the same features of conventional exciton-polariton condensation in the system ground state, verify the non-zero momentum condensation near these Dirac points

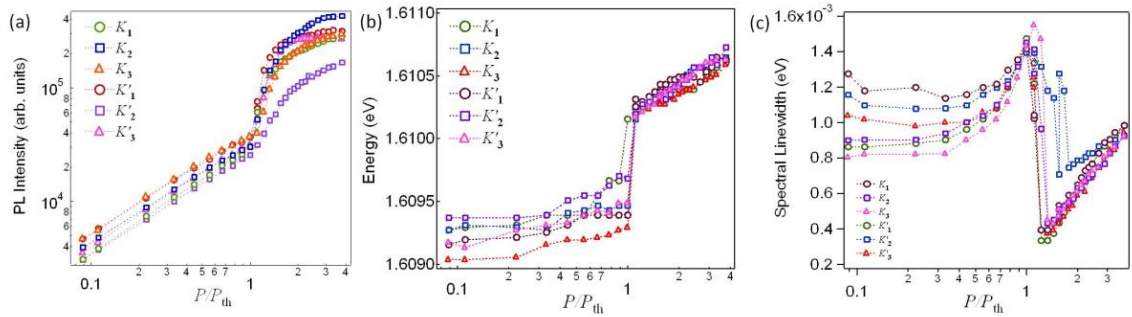

**Figure 3** Exciton-polariton condensates in Dirac points. (a) The input-output characteristic in the log-log scale at all three $K$ and three $K'$ points. The unequivocal nonlinear intensity increase appear above $P/P_{th} > 1$. Condensate energy (b) and spectral linewidth (c) are plotted against $P/P_{th}$.

3.2.2 Experimentally constructed band structures and Dirac dispersions

The tomography of energy-resolved momentum space spectroscopy allows us to construct the energy band structures of LPs in weak triangular lattice potentials. Figures 4(a) through 4(d) show the evolution of the constructed band structures in the reduced zone scheme as a function of the laser pump powers, $P/P_{th} \sim$ 0.2, 0.9, 1.3 and 5. Below threshold pump powers ($P/P_{th} < 1$), the band structures are simply the parabolic dispersions of thermal LPs, from which the effective mass of LPs is estimated from the parabola energy dispersion curvature. Lack of coherence among thermal LPs together with broad spectral linewidth masks any visible effect of the weak triangular potential in the band structure. As the density of LPs reaches near the quantum degeneracy point at $P/P_{th}$ ~ 0.9 (Fig. 4(b)), the dispersions become broader arising from the increased polariton-polariton scatterings, and LPs start being accumulated at local meta-stable $K$ and $M$ points of the band structures. This feature matches with the formation of the first hexagonal BZ in the LP momentum distributions (Fig. 2(b), where the sharp Bragg peaks at $K$ and $K'$ points are seen. When LPs are further injected above threshold, the bands become sharper i.e. the reduced spectral linewidth (Fig. 3(c)), the lower energy bands at $\Gamma$ point exhibit a blue energy shift by ~ 1.4 meV, and the gapped higher bands at $K$ and $M$ points become visible (Fig. 4(c)). The estimated energy gap between the first and second/third bands at $K$ points is ~ 86 $\mu$eV, which is compatible to the value of the potential strength (~ 200 $\mu$eV). At $P/P_{th}$ ~ 5, the band structures at $K$ points smear, which would result from scatterings into $\Gamma$ point, and the intensity of LPs grows at $\Gamma$ point.

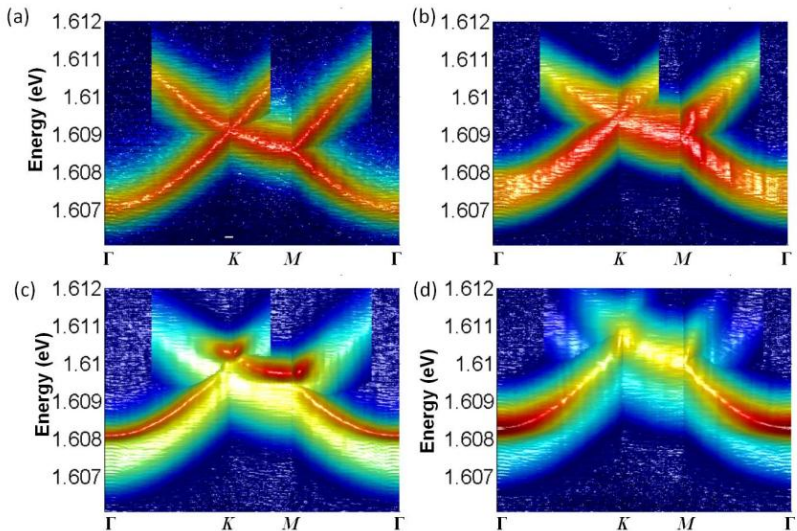

**Figure 4** Experimentally taken band structures of lower polaritons at $P/P_{th}$ ~ 0.2 (a), 0.9 (b), 1.3 (c), and 5 (d) along three high symmetry points $\Gamma$, $M$, $K$ defined in Fig. 1(b). The higher (lower) intensity is presented in color of red (blue).

Exciton-polariton condensates near the *K* points in the range of 1< $P/P_{th}$ < 5 would be stabilized by a bottleneck effect as a competition of the finite life time and relaxation time. Although we are unable to resolve the crossing feature of the second and third bands at Dirac points partially due to ~ 250 $\mu$eV-broad LP condensate signals, we identify the first and third bands in the **Γ- *K*** zone, the second and third bands in the ***K-M*** zone and the first and second bands in the ***M*-Γ** zone. We clearly capture the signatures of strong accumulated LPs near *K* points and the linear dispersions near *K* points, consistent with theoretical single-particle band structure result. More closely, our experimental data is fitted with the computed band structures introducing the only one fitting parameter, the characteristic energy $E_0$. The periodic potential strength $V_0$ is fixed at 0.05 $E_0$. Figure 5(a) is a zoomed-in view of our attempt on the data at $P/P_{th}$ ~ 1.3 with $E_0$ = 4.5 meV, which is given by the measured effective mass below threshold and the lithographic lattice constant $a = 2$ $\mu$m. The full fitting results at different pump values are given along the horizontal direction through $K_1$ point in Fig. 6 and along the vertical direction through $K_1$ point in Fig. 7. LPs are accumulated at *K* and *K'* points from the higher bands through several paths. For examples, LPs at **Γ** point of the third band will fall into *K* directly from **Γ** to *K* (**Γ-*K*** path, green arrow (Fig. 1(b))) or from **Γ** via *M* to *K* (**Γ-*M-K*** path, green arrows (Fig. 1(b))) inside the third bands. The Dirac points are protected by a gap in the **Γ-*M-K*** path, meanwhile some of LPs will be leaked out to *M* from the LP **Γ- *K*** path as seen in Fig. 4(b). In addition, the stability of polaritons near the *K* points would be related to superfluidity fingerprints predicted for interacting bosons. However, we are not yet reaching any clear superfluid features in the band structures partially due to the finite linewidth of LP signals in this current sample.

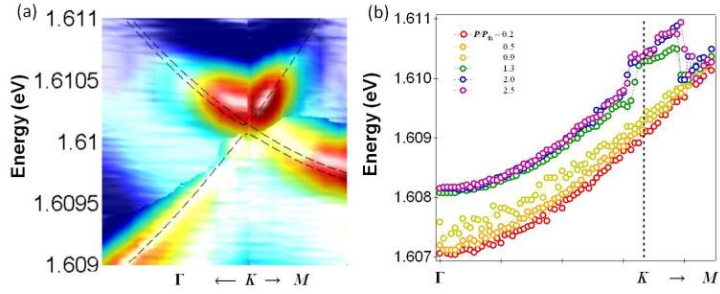

**Figure 5** (a) Energy dispersions at $P/P_{th} \sim 1.3$ along the horizontal arrow at $K_1$ denoted in Fig. 1(c). The dashed black lines are the first three Bloch bands computed in a single particle plane-wave basis with the parameters, $E_0 = 4.5$ meV, $V_0 = 0.05\,E_0$, and the energy offset 1.4 meV at $\Gamma$ for the energy blue-shift due to the polariton-polariton interaction. (b) The extracted energy values at the maximum LP emission intensity are plotted against the loop of $\Gamma$, $M$ and $K$. We have calculated the Dirac polariton velocity from the linear regression analysis from the $K$-$M$ section.

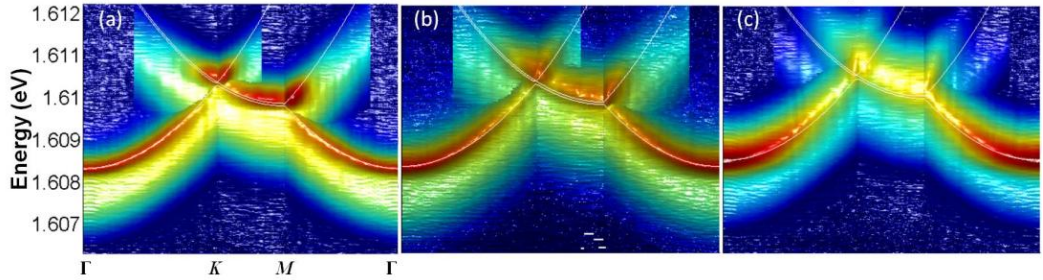

**Figure 6** Experimentally constructed band structures fitted with the first three theoretical band structures (white lines) along the whole loop of high symmetry points from at $P/P_{th} \sim$ 1.3 (a), 2 (b), and 5 (c). The only fitting parameter is the characteristic energy $E_0 \sim 4.5$ meV, and the energy blueshift has been taken into account as a linear shift by 1.05 meV, 1.1 meV, 1.25 meV at $P/P_{th} \sim 1.3$, 2 and 5, respectively.

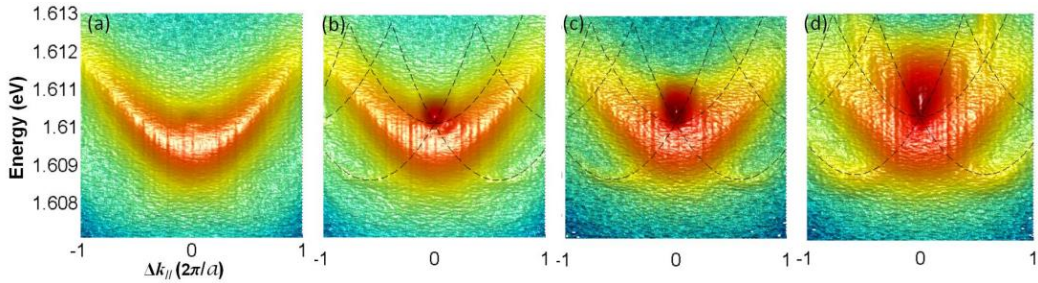

**Figure 7** Experimentally constructed band structures fitted with theoretically computed band structures (Black dashed lines) along the vertical direction through $K_1$ (Fig. 1c) at $P/P_{th} \sim 0.9$ (a), 1.3 (b), 2 (c), and 5 (d). Two fitting parameters, the characteristic energy $E_0$ and the energy blue-shift have the same values which are used for the fitting in Fig. 6.

Quantifying the slope of the linear Dirac dispersion, we perform the linear regression analysis along two orthogonal directions around $K_1$ point indicated with two black arrows in Fig. 1(c). Figure 5(b) is the maximum intensity plot of the cross-sectional cut in the horizontal direction from $\Gamma$ through $K_1$ to $M$ at varying pump power values, $P/P_{th}$ ~ 0.2, 0.5, 0.9, 1.3, 2.0 and 2.5. The extracted velocity values are $1.0\times10^8$ cm/s for $P/P_{th}$ ~ 1.3 and $2.1\times10^8$ cm/s, almost 2 times faster for a higher pumping rate $P/P_{th}$ ~ 2. Both values are in the same order with the predicted velocity value, $1.5\times10^8$ cm/s for given LP effective mass $m^*$ and the lattice constant a values. We conjecture that the steeper slope at higher pump rates would be associated with the increased repulsive interactions among particles. The quantitative study of the interaction effect deserves to receive full attention, leaving for future investigation. At high above threshold values, the reliable linear regression analysis fails, where the relaxation to the $\Gamma$ point become dominant as seen in Fig. 4(d).

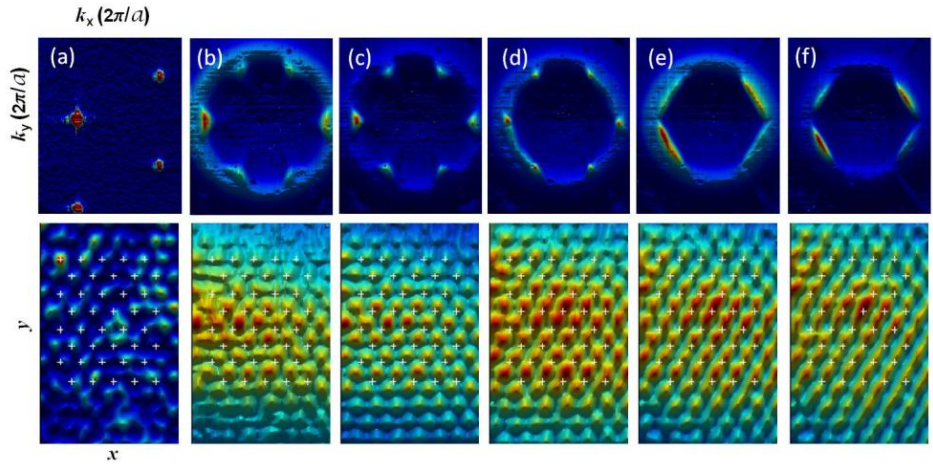

**Figure 8** Equi-energy maps in the momentum space $(k_{//,x}, k_{//,y})$ (top panel) and in the real space $(x, y)$ (bottom panel). The distributions in (a) are at the pump laser energy value 1.6161 eV, and the distributions in (d) are at the Dirac point energy, 1.6102 eV. The maps in (b) and (c) are the states whose energy values are 532 $\mu$eV and 353 $\mu$eV respectively above the Dirac point energy (d). The plots in (e) and (f) are the state below -175 $\mu$eV and -353 $\mu$eV from the Dirac point energy (D), respectively. The crosses in the bottom panel are located at the position of the patterned apertures. In both panels, red color represents high intensity and blue does low intensity.

Lastly, the position-resolved spectroscopy measurements are executed for obtaining the LP population distribution in the real space, which would be related with the LP population distribution in momentum space via Fourier transformation. The top panel in Fig. 8 displays the equi-energy map at different energy values in the momentum space taken at $P/P_{th}$ ~ 1.3, and the associated near-field LP intensities are plotted in the bottom panel. Figures 8 (a) are from the scattered pump laser light, from which the positions of the lithographically patterned traps are indicated with crosses in the bottom panel since the laser lights are scattered from the metal film of the traps. The distance between the nearest neighbour crosses is $(2 \pm 0.27)\ \mu m$, which is the device lattice constant. First of all, these equi-energy maps in the momentum space enable us to trace how LPs are relaxed from higher to lower energy states. For the states at Dirac points and near them, the clear triangular patterns of the strong intensity peaks in the real space appear outside not at the trap positions. This observation means that the orbital symmetry of a macroscopic order parameter at this energy state is $p$-like wave, which is responsible for the second and third bands around Dirac points, which are consistent with the single particle band structure theoretical results.

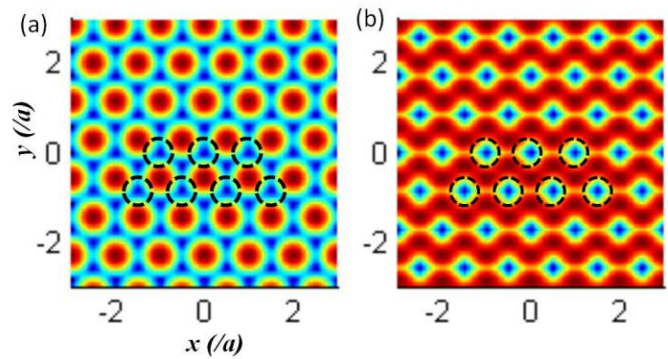

**Figure 9** Intensity plots of coherently superposed two degenerate $p$-like states(A) and statistical mixture of two degnerate $p$-like states (B) both at $K$ and $K'$ points. Black circles are at the trap positions. Red and blue colors in the color images represent high and low intensity values.

In order to explain the position-resolved equi-energy map in the real space (Fig. 8 bottom panel), we look at individual wavefunction intensity patterns of degenerate doublet states at $K$ and $K'$ points obtained from the band structure calculation. The individual $p$-like wavefunctions do not form the triangular pattern observed in experiments (Fig. 8(d), bottom). Two possible candidates for such triangular peak patterns are plotted in Fig. 9, consisting of two degenerate $p$-like wavefunctions $p_1$, $p_2$: (a) the coherent superposition, $|p_1 + ip_2|^2$ resembling the vortex states and (b) the statistical mixture, $|p_1|^2+|p_2|^2$. Since the intensity of $p$-like wavefunctions is zero at the core, the minimum (blue) appear at the center of traps. Although the triangular patterns appear in both cases, the coherent superposition yields the peak-array triangular pattern, whereas the statistical mixture does the valley-array triangular pattern. Our observed wavefunctions in the real space (Fig. 8(d) bottom) may resemble Fig. 9(a), which imply that the states at $K$ and $K'$ points would form vortex states. However, the unequivocal evidence for the vortex state formation at Dirac points should be acquired carefully via interferometers, where essential phase information in space leads a decisive conclusion to distinguish coherent vortex states from the statistical mixtures of two degenerate $p$-like states.

4. Conclusion

We have observed exciton-polariton condensates near the Dirac points formed at high symmetry $K$ and $K'$ points of a 2D triangular lattice, originating from the two-degenerate $p$-like orbitals. Our photoluminescence spectroscopy and images in the momentum and the real spaces allow us to directly measure the Dirac relations and their corresponding orbital symmetries together. Due to the dynamical nature of exciton-polariton condensates, we are able to directly access the states near the Dirac points, which would be often harder to approach because of complexities in condensed matter systems. Furthermore, when the degenerate states are lifted and the time-reversal symmetry is broken, the states near Dirac points are predicted to propagate unidirectionally, which would be interpreted to be the analog of electron's chiral edge states in the quantum Hall regime [16-18]. In this regime, interesting quantum order states, for example, appearing in graphene and topological insulators can be explored in the polariton-lattice system both in the real and reciprocal spaces as well as nonlinear interactions among particles.

**Acknowledgements**: This work has been supported by Navy/SPAWAR Grant N66001-09-1-2024, the Japan Society for the Promotion of Science (JSPS) through its "Funding Program for World-Leading Innovative R&D on Science and Technology (FIRST Program)", and the State of Bavaria. N. Y. K appreciates C. Wu, Z. Cai, K. Fang, and P. Kim for insightful discussion.